\documentclass[12pt]{article}
\usepackage{epsfig,graphicx}
\usepackage{fpcp03}

\newcommand{\nn}{\nonumber} 
\newcommand{\bea}{\begin{eqnarray}}
\newcommand{\eea}{\end{eqnarray}}
\newcommand{\mcdot}{\!\cdot\!}
\newcommand{\bn}{\bar n}

\def\dslash{\partial\!\!\!\slash}
\def\Dslash{D\!\!\!\!\slash}

\def\bnslash{\bar n\!\!\!\slash}

\newcommand{\SCETa}{\mbox{${\rm SCET}_{\rm I}$ }}
\newcommand{\SCETb}{\mbox{${\rm SCET}_{\rm II}$ }}

\def\beq{\begin{equation}}
\def\eeq#1{\label{#1}\end{equation}}
\def\eeqn{\end{equation}}

\def\beqa{\begin{eqnarray}}
\def\eeqa#1{\label{#1}\end{eqnarray}}
\def\eeqan{\end{eqnarray}}

\let\bar=\overbar

\def\Dslash{\not{\hbox{\kern-4pt $D$}}}
\def\dslash{\not{\hbox{\kern-2pt $\del$}}}

\def\msb{{\bar{\ssstyle M \kern -1pt S}}}

\def\BB0bar{B^0 {\overline B}^0}
\def\BB0dbar{B_d^0 {\overline B}_d^0}
\def\BB0sbar{B_s^0 {\overline B}_s^0}

\RequirePackage{xspace}

\usepackage{relsize}
\def\babar{\mbox{\slshape B\kern-0.1em{\smaller A}\kern-0.1em
    B\kern-0.1em{\smaller A\kern-0.2em R}}}

\def\Kbar  {\kern 0.2em\overline{\kern -0.2em K}{}\xspace}

\def\Kz    {\ensuremath{K^0}\xspace}
\def\Kzb   {\ensuremath{\Kbar^0}\xspace}
\def\KzKzb {\ensuremath{\Kz \kern -0.16em \Kzb}\xspace}
\def\Kp    {\ensuremath{K^+}\xspace}
\def\Km    {\ensuremath{K^-}\xspace}

\def\KpKm  {\ensuremath{\Kp \kern -0.16em \Km}\xspace}

\def\Dbar    {\kern 0.2em\overline{\kern -0.2em D}{}\xspace}

\def\Dz      {\ensuremath{D^0}\xspace}
\def\Dzb     {\ensuremath{\Dbar^0}\xspace}
\def\DzDzb   {\ensuremath{\Dz {\kern -0.16em \Dzb}}\xspace}
\def\Dp      {\ensuremath{D^+}\xspace}
\def\Dm      {\ensuremath{D^-}\xspace}

\def\DpDm    {\ensuremath{\Dp {\kern -0.16em \Dm}}\xspace}

\def\Bbar    {\kern 0.18em\overline{\kern -0.18em B}{}\xspace}

\def\BB      {\ensuremath{B\Bbar}\xspace} 
\def\Bz      {\ensuremath{B^0}\xspace}
\def\Bzb     {\ensuremath{\Bbar^0}\xspace}
\def\BzBzb   {\ensuremath{\Bz {\kern -0.16em \Bzb}}\xspace}
\def\Bu      {\ensuremath{B^+}\xspace}
\def\Bub     {\ensuremath{B^-}\xspace}

\def\BpBm    {\ensuremath{\Bu {\kern -0.16em \Bub}}\xspace}

\mathchardef\Upsilon="7107
\def\Y#1S{\ensuremath{\Upsilon{(#1S)}}\xspace}

\mathchardef\Deltares="7101
\mathchardef\Xi="7104
\mathchardef\Lambda="7103
\mathchardef\Sigma="7106
\mathchardef\Omega="710A

\def\Deltabar{\kern 0.25em\overline{\kern -0.25em \Deltares}{}\xspace}
\def\Lbar{\kern 0.2em\overline{\kern -0.2em\Lambda\kern 0.05em}\kern-0.05em{}\xspace}
\def\Sigbar{\kern 0.2em\overline{\kern -0.2em \Sigma}{}\xspace}
\def\Xibar{\kern 0.2em\overline{\kern -0.2em \Xi}{}\xspace}
\def\Obar{\kern 0.2em\overline{\kern -0.2em \Omega}{}\xspace}
\def\Nbar{\kern 0.2em\overline{\kern -0.2em N}{}\xspace}
\def\Xb{\kern 0.2em\overline{\kern -0.2em X}{}\xspace}

\newcommand{\tev}{\ensuremath{\mathrm{\,Te\kern -0.1em V}}\xspace}
\newcommand{\gev}{\ensuremath{\mathrm{\,Ge\kern -0.1em V}}\xspace}
\newcommand{\mev}{\ensuremath{\mathrm{\,Me\kern -0.1em V}}\xspace}
\newcommand{\kev}{\ensuremath{\mathrm{\,ke\kern -0.1em V}}\xspace}
\newcommand{\ev}{\ensuremath{\mathrm{\,e\kern -0.1em V}}\xspace}
\newcommand{\gevc}{\ensuremath{{\mathrm{\,Ge\kern -0.1em V\!/}c}}\xspace}
\newcommand{\mevc}{\ensuremath{{\mathrm{\,Me\kern -0.1em V\!/}c}}\xspace}
\newcommand{\gevcc}{\ensuremath{{\mathrm{\,Ge\kern -0.1em V\!/}c^2}}\xspace}
\newcommand{\mevcc}{\ensuremath{{\mathrm{\,Me\kern -0.1em V\!/}c^2}}\xspace}

\def\mus  {\ensuremath{\rm \,\mus}\xspace}

\def\mus        {\ensuremath{\,\mu{\rm s}}\xspace}    

\def\to                 {\ensuremath{\rightarrow}\xspace}

\def\pep2{PEP-II}

\def\gsim{{~\raise.15em\hbox{$>$}\kern-.85em
          \lower.35em\hbox{$\sim$}~}\xspace}
\def\lsim{{~\raise.15em\hbox{$<$}\kern-.85em
          \lower.35em\hbox{$\sim$}~}\xspace}

\xspace

\def\jetset74   {\mbox{\tt Jetset \hspace{-0.5em}7.\hspace{-0.2em}4}\xspace}

\begin{document}


\begin{flushright}
MIT-CTP 3411
\end{flushright}

\Title{The phenomenology of rare and semileptonic B
decays\footnote{Invited talk given by D.P. at the 2$^{\rm nd}$ Conference on
Flavor Physics and CP Violation, FPCP 2003, 3-6 June 2003, Paris, France.} }
\bigskip


%
\label{PirjolStart}

%
\author{ Dan Pirjol $^{a,b}$ and Iain W. Stewart $^b$}

%
\address{$^a$Department of Physics and Astrophysics\\
The Johns Hopkins University\\
Baltimore MD 21218 \\
$^b$Center for Theoretical Physics\\
Massachusetts Institute of Technology\\
Cambridge, MA 02139 \\
}

\makeauthor\abstracts{ We summarize recent progress in the theory of exclusive
  rare and semileptonic B decays, focusing on model-independent results. The
  heavy-to-light form factors parameterizing these decays admit a
  model-independent description in two distinct kinematical regions.  In the
  large-energy limit of an energetic light meson, the Soft-Collinear Effective
  Theory can be used to prove factorization formulas for the form factors. We
  present factorization formulas for all $B\to P, V$ form factors at leading
  order in $\Lambda/m_b$.  Near the zero-recoil point, Heavy Quark Effective
  Theory gives useful relations among the form factors of different currents.  }

\section{Introduction}

The exclusive B decays are, in many ways, unique probes of the 
Standard Model and its extensions. 
The semileptonic $B$ decays to charmless states can give
information about $|V_{ub}|$, while the exclusive radiative decays
$B\to \rho(K^*)\gamma$ decays can be used to extract $|V_{td}|$.
In addition, many of these decays are flavor changing neutral current
processes, which proceed only through loops. Therefore they
are sensitive to the presence of new nonstandard particles running
in the loop, and can be used to search for new physics effects \cite{rare}.

Experimental data on these decays is becoming available, including
not only branching ratios, but also spectrum shapes in semileptonic
decays \cite{exp}.
Interpreting this data for the purpose of extracting CKM parameters and  
in searching for New Physics effects requires good control over the 
Standard Model description of these decays. Many computations of these
form factors are available, using methods as varied as quark models,
QCD sum rules (see \cite{pball} for a recent review) and lattice QCD 
\cite{becirevic-fpcp03}. We will focus here on recent model independent 
results.

There are two kinematical regions where model independent results can 
be established, corresponding to the two kinematical limits of: a) slow and
b) energetic final light hadrons. They are most naturally discussed in terms
of two effective theories: a) Heavy Quark Effective Theory (HQET) and
b) Soft-Collinear Effective Theory (SCET). 
Rather than following the historical order of events, we will discuss these
two types of predictions starting from the respective effective theory describing
each of these two situations.
The large energy limit will be discussed in Sec.~II and the case of the
low recoil form factors is covered in Sec.~III. An Appendix contains a summary of
the factorization formulas for the $B\to P,V $ form factors contributing
to rare and semileptonic $B$ decays.

\section{Large energy limit and the SCET}

The Soft-Collinear Effective Theory (SCET) was proposed in \cite{scet}
as a systematic framework for the study of processes involving energetic
quarks and gluons, and was discussed in detail in another talk at this
Conference \cite{fleming-fpcp03}. In this talk we summarize only a 
few points strictly necessary for the discussion of the form factors.

The SCET separates the contributions
from the different energy scales relevant in a physical problem.
This is done by introducing fields with well-defined momentum scaling,
corresponding to the modes relevant for reproducing the IR of the
full theory. These modes include i) collinear quarks $\xi_n$ and
gluons $A_n$ with momenta $p_c \sim Q(\lambda^2,1,\lambda)$,
ii) usoft quarks $q$ and gluons $A_\mu$ with momenta $p_{us} \sim
Q\lambda^2$ and iii) soft modes $q_s, A_s^\mu$ with momenta $p_s\sim
Q\lambda$. The definition of the expansion parameter $\lambda$ depends
on the specific problem being studied. We use here and below light-cone
component notation $p = (n\cdot p, \bn\cdot p, p_\perp)$, defined in 
terms of light cone unit vectors $n^2 = \bn^2 = 0, n\cdot \bn = 2$.

The couplings of the effective theory fields are described by the
SCET Lagrangian. In a theory containing only usoft and collinear modes,
these couplings can be written as\footnote{The complete case containing
also soft fields is discussed in Refs.~\cite{bpssoft,HiNe,bps5,gauge}.}
\bea
{\cal L}_{SCET} = {\cal L}_{\xi\xi} + {\cal L}_{cg} + {\cal L}_{q\xi}\,,
\eea
where the first two terms describe couplings of collinear fields to 
each other
$({\cal L}_{\xi, cg})$ \cite{scet} and usoft-collinear interactions 
$({\cal L}_{q\xi})$ \cite{bcdf,ps1}, respectively.
They can be expanded in $\lambda$ as
\bea\label{Lqxi}
{\cal L}_{\xi\xi} = {\cal L}^{(0)}_{\xi\xi} + 
{\cal L}^{(1)}_{\xi\xi} + \cdots\,,\qquad\qquad
{\cal L}_{q\xi} = {\cal L}^{(1)}_{q\xi} + {\cal L}^{(2)}_{q\xi} + \cdots\,,
\eea
where the leading order collinear quark Lagrangian is (with $iD^\mu_{\rm us} =
i\partial^\mu + g A^\mu_{\rm us}$)
\bea\label{Lxi0}
{\cal L}^{(0)}_{\xi\xi} = \bar \xi_n \left\{ n\cdot iD_{\rm us} + 
gn\cdot A_n +
i\Dslash_{\perp c} \frac{1}{\bn\cdot iD_c} i\Dslash_{\perp c} \right\}
\frac{\bnslash}{2}\xi_n
\eea
The explicit form of ${\cal L}_{cg}$ can be found in Ref.~\cite{bpssoft}.
Note the fact that the usoft-collinear Lagrangian ${\cal L}_{q\xi}$ starts
at subleading order with terms of $O(\lambda)$. 
The weak current $\bar q\Gamma b$ is analogously matched onto SCET operators
as \cite{scet,rpi,bcdf,ps1}
\bea\label{J}
\bar q\Gamma b = \int \mbox{d}\omega C_0(\omega) J_0(\omega) + 
\int \mbox{d}\omega C_{1a}(\omega) J_{1a}(\omega) + 
\int \mbox{d}\omega_1 \mbox{d}\omega_2 
C_{1b}(\omega_1,\omega_2) J_{1b}(\omega_1,\omega_2) + \cdots
\eea
where the ellipses denote operators suppressed by $O(\lambda^2)$. The most general 
form of these
operators is given in Ref.~\cite{ps1} for all allowed Dirac structures $\Gamma$.

An important property of the effective theory is ultrasoft-collinear
factorization at leading order in $\lambda$. Since the usoft gluons couple
to collinears only through the first term in Ref.~(\ref{Lxi0}), their
effects can be absorbed at this order into a Wilson line $Y_n[n\cdot A]$
by performing a field redefinition of the collinear fields \cite{bpssoft}
\bea\label{redef}
\xi_n = Y_n[n\mcdot A_{\rm us}] \xi_n^{(0)}\,,\quad
A_n^\mu = Y_n A_n^{(0)\mu} Y^\dagger_n \,,\quad
Y_n[n\mcdot A] \equiv P\exp\left(ig \int_{-\infty}^0 ds n\mcdot A_{\rm us}(ns)\right)\,.
\eea
The new collinear fields $\xi_n^{(0)}$ and $A_n^{(0)}$ do not couple
to the usoft gluon field $A_{\rm us}$, which now appears only through the Wilson
line $Y[n\cdot A_{\rm us}]$.

\subsection{Factorization in $B\to \gamma \ell \nu$}

The simplest exclusive heavy meson process which can be described in this
framework is the leptonic radiative decay $B^\pm \to \gamma \ell^\pm \nu$. 
This decay proceeds through weak annihilation of the $B$ constituent quarks
and does not suffer from the chirality suppression affecting the pure leptonic 
decay $B\to \ell \nu$. Model dependent estimates \cite{b2gpapers} suggest
branching ratios for this mode of the order of $\sim 10^{-6}$, which should
be within the reach of the B factories.

We will be interested in the kinematical region where the photon energy
$E_\gamma$ is much larger than $\Lambda_{QCD}$, but can be smaller or comparable
with the heavy quark mass $m_b$. In this region there are three relevant
energy scales: the hard scale $Q$, with $Q=\{m_b, E_\gamma\}$, the collinear
scale $p_c^2 \sim Q\Lambda$, and the soft scale $p_s^2 \sim \Lambda^2$.
The soft scale is introduced by the typical momenta of the soft spectators
in the $B$ meson, and the collinear scale gives the typical virtuality of
a spectator quark after being struck by the energetic photon $p_c = p_{sp} + q$.
Finally, the hard scale is associated with hard gluons with virtualities of 
the order of the heavy quark mass.

Using SCET methods, a factorization theorem was proved in \cite{lpw} to
all orders in $\alpha_s$, expressing the form factors for this mode at leading
order in $\Lambda/Q$ as (with $Q=\{m_b, E_\gamma\}$)
\bea\label{b2g}
f_{V,A}(E_\gamma) = \frac{Q_q f_B m_B}{2E_\gamma} C_{V,A}(E_\gamma,\mu)
\int \mbox{d}k_+ \frac{1}{k_+} J(E_\gamma k_+,\mu) \phi_B^+(k_+,\mu)
\eea
The three factors in this formula are connected with the three
different scales in this problem: the Wilson coefficients $C_{V,A}$
appear in the matching of the heavy-to-light currents $\bar u \gamma_\mu (\gamma_5) b$
onto SCET operators, the jet function $J(E_\gamma k_+,\mu) = 
1 + O(\alpha_s(p_c^2))$ accounts for effects associated with the collinear scale,
and the B meson light-cone wave function (normalized as $\int dk_+ \phi_B^+(k_+) = 1$)
accounts for fluctuations over the scale of the soft modes.
Factorization in $B\to \gamma e\nu$ was also studied in \cite{b2gpapers,DS}.

We mention here a few implications of the factorization formula Eq.~(\ref{b2g}).
At tree level in matching at the scale $Q\Lambda$, it predicts that the form factors in
$B\to \gamma \ell\nu$ are proportional to the first inverse momentum of the
$B$ light-cone wave function $\langle k_+^{-1} \rangle$. The same moment appears 
in many other factorization
formulas for B meson decays. Therefore, measurements of the
photon spectrum in $B\to \gamma \ell\nu$ could provide a model-independent
extraction of this parameter.
Second, all LO form factors determining $B\to \gamma \ell \nu$, 
$B_s\to \gamma \ell^+ \ell^-$ and $B\to \gamma\gamma$ decays are given
by one single nonperturbative integral over the $B$ wave functions \cite{lpw,DS}.
Therefore their ratios can be computed in terms of the Wilson coefficients
$C_{V,A,T}$ which have expansions in $\alpha_s(Q)$ and contain
double Sudakov logs. 
Finally, the corrections to the factorization formula Eq.~(\ref{b2g}) are 
suppressed by $\Lambda/Q$ and come from matrix elements of power suppressed 
operators in the SCET.

\subsection{Factorization for heavy-light form factors}

We consider next the case of the heavy-to-light form factor, relevant for 
the semileptonic decays $B\to \pi (\rho ) \ell \nu$, or the rare radiative decays
$B\to K^*\gamma, K^* \ell^+\ell^-$. The dynamics for this case is more
complicated than for the leptonic radiative decay due to the presence of 
the collinear partons in the final state light meson.

\begin{figure}[!t]
 \centerline{
  \mbox{\epsfxsize=5.0truecm \hbox{\epsfbox{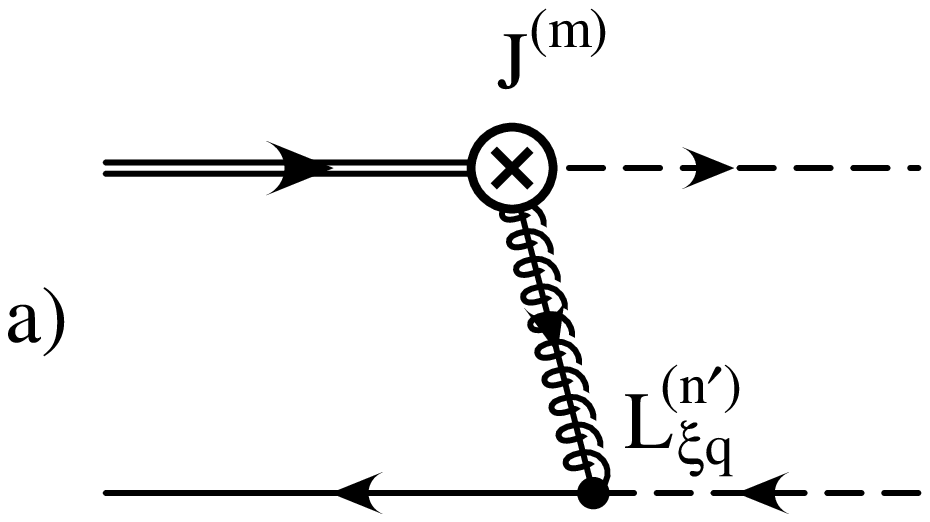}} }\hspace{1.0cm}
  \mbox{\epsfxsize=5.0truecm \hbox{\epsfbox{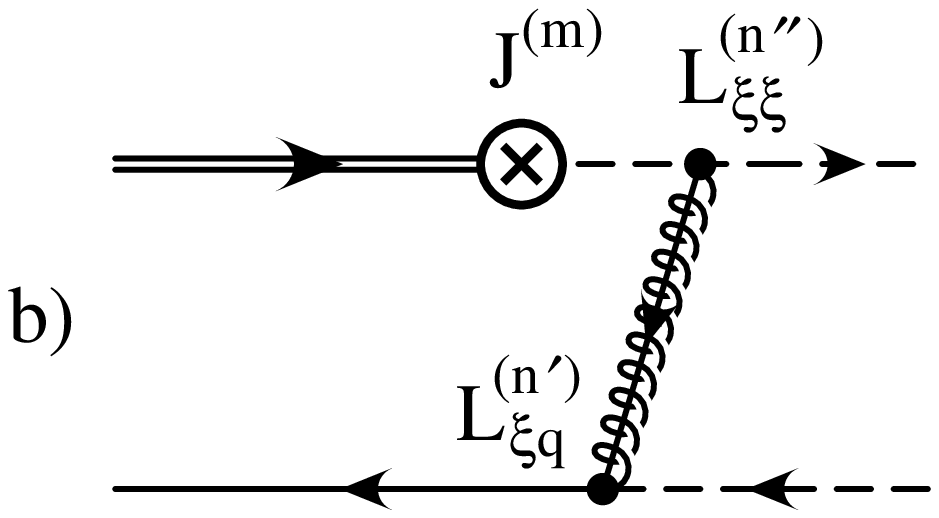}} }
  } 
\caption[1]{Tree level contributions to the $T$-products (\ref{Tproducts}) in 
\SCETa.}
\label{fig_eft} 
\end{figure}

Although it had been known for a long time that exclusive hadronic form factors
admit a systematic expansion in perturbative QCD at large momentum transfer
$Q$ \cite{BL}, the extension of this approach beyond leading order in 
$\Lambda/Q$ has been a difficult task. At subleading order one
quickly encounters difficulties connected with soft parton configurations,
corresponding to situations where one of the partons in a given hadron
carries most of the hadron momentum. In heavy-to-light problems, such 
effects appear already at leading order in $\Lambda/m_b$.
In the standard hard scattering analysis, they 
lead to unintegrable singularities in the hard scattering 
kernels \cite{akhoury} which at $O(\alpha_s)$ can be absorbed into
soft form factors $\zeta$ \cite{BeFe}.

The heavy-light form
factor was recently studied in \cite{bps5} using the SCET, where a factorization 
theorem at leading order in $\Lambda/Q$ was established.
The main points are:

\begin{itemize}

\item There are two relevant perturbative scales in this problem: 
$Q$ and $\sqrt{\Lambda Q}$, where $Q = \{m_b, E_\pi\}$. The 
effects associated with these two
scales can be included using a two-step matching: QCD $\to$ \SCETa $\to$
\SCETb. Here \SCETa
contains collinear modes with $p_c^2 \sim Q\Lambda$ and usoft
modes with $p_{us}^2 \sim \Lambda^2$, and \SCETb
includes collinear and soft modes with $p_c^2 = p_s^2 = \Lambda^2$.

\item The light meson state is purely collinear, and couples to the
soft $B$ meson state through the weak currents $J$ given in Eq.~(\ref{J}), 
and the usoft-collinear Lagrangian ${\cal L}_{q\xi}$ in Eq.~(\ref{Lqxi}).
In \SCETa, the weak current contributing to 
the heavy-to-light form factor is matched at LO onto the $T$-products
\begin{eqnarray} \label{Tproducts}
T_1^{\rm F} \!\!&=&\!\! T \big[J^{(1a)}, i{\cal L}^{(1)}_{\xi q} \big],\
 \qquad
 T_2^{\rm F} =\! T \big[J^{(1b)}, i{\cal L}^{(1)}_{\xi q} \big] \,, \\
T^{NF} \!\!&=&\!\!  T \big[J^{(0)}, i{\cal L}^{(1)}_{\xi\xi},
  i{\cal L}^{(1)}_{\xi q} \big] + T \big[J^{(0)}, i{\cal L}^{(2)}_{\xi q} \big] +
 T \big[J^{(0)}, i{\cal L}^{(1)}_{cg},i{\cal L}^{(1)}_{\xi q} 
  \big] + T \big[J^{(0)}, i{\cal L}^{(1)}_{\xi q} \big] \,. \nn \hspace{0.cm}
\end{eqnarray}
These contributions are shown in a graphical form in Fig.~\ref{fig_eft}.

\item After usoft-collinear factorization Eq.~(\ref{redef}), the T-products
$T_{1,2}^F$ factor, and can be matched directly onto \SCETb operators.
Their matrix elements give the `factorizable' contribution. On the other
hand, the $NF$ $T-$products all involve $J_0$, and their matrix elements are 
parameterized in terms of three `soft' form factors, appearing in the same 
combinations as derived in \cite{Charles}.

\end{itemize}

In this way one finds that at leading order in $\Lambda/Q$, 
the heavy-to-light $B\to M$ form factors 
can be represented by a factorization formula  written schematically as
\bea\label{fact}
f_i(q^2) = C_{ij}(Q) \zeta^M_j(Q\Lambda,\Lambda^2) + 
\int dx dz dk_+ C_{ij}(z,Q) 
J_j(z,x,k_+)  \phi_B^+(k_+)  \phi^M_j(x)
\eea
Here $C_{ij}$ are Wilson coefficients of \SCETa operators, and $J_j(z,x,k_+)$
are jet functions appearing as Wilson coefficients in matching onto \SCETb operators. 
Both these
quantities are computable in perturbation theory. The nonperturbative
effects in Eq.~(\ref{fact}) occur in the form factors $\zeta_i$,
and the light-cone wave functions of the $B$ meson and
of the light meson $\phi_B^+(k_+), \phi_i(x)$. In the Appendix we give
the explicit 
form of the factorization formulas for all $B\to P,V$ form factors 
relevant for phenomenological applications.

A factorization formula for the heavy-light form factor of the form
$f = C\zeta + \phi_B \otimes T\otimes \phi_M$
was first proposed in 
Ref.~\cite{BeFe}, based on the study of low order contributions in
perturbation theory. The effective theory analysis in \cite{bps5} establishes
such a factorization theorem in the form of Eq.~(\ref{fact}) to all orders in 
$\alpha_s$, and at leading order in $\Lambda/Q$. It also shows that $\zeta$
depends on the scale $Q\Lambda$, as well as $\Lambda^2$.

The two terms in the factorization formula Eq.~(\ref{fact})
are of the same order in $\lambda=\Lambda/Q$, such that their relative numerical
contributions can be comparable. Their scaling can be obtained from a
simple model independent power counting argument as follows \cite{bps5}. 
The \SCETa operators in 
Eq.~(\ref{Tproducts}) scale like $\lambda^3$. After matching onto \SCETb
the scaling of the collinear fields gets an additional $\lambda$.
This gives for the total scaling of each of the two terms in Eq.~(\ref{fact})
$\lambda^3 \times \lambda \times \lambda^{-1}\times
\lambda^{-3/2} \sim \Lambda^{3/2}$, where the factors of $\lambda^{-1}$ and
$\lambda^{-3/2}$ correspond to the scaling of the light meson and B states,
respectively.

Although the numerical values of the soft form factors $\zeta$ are not constrained
by the effective theory, and have to be obtained from model computations or lattice
QCD, the factorization results have significant predictive power. For example,
using as input the form factor $f_+(q^2)$ as measured in $B\to \pi e\nu$, the 
remaining $B\to \pi$ form factors can be computed using 
Eqs.~(\ref{f+})-(\ref{fT}) and $\phi_B(k_+),\phi_\pi(x)$.
Finally, the explicit results in Eqs.~(\ref{f+})-(\ref{T1}) can be used to
calculate Sudakov effects from the RG running of the Wilson coefficients $C_i, B_i$.
It will be interesting to see how the results of this running compare with the 
results in Refs.~\cite{akhoury,sudakov,li,lireview}.

The factorization relations have to be extended for the case of the penguin
mediated rare radiative
decays such as $B\to V\gamma$ and $B\to V \ell^+ \ell^-$ ($V=K^*,\rho$), to account
for the contributions of weak 4-quark operators. These effects have been computed in
Refs.~\cite{4q,wa} and contribute about 5-10\% to the observed branching ratios.

\section{Zero recoil region}

In the low recoil region for the final meson, corresponding to
maximal $q^2 \sim (m_B-m_M)^2$, heavy quark symmetry can be applied to describe 
the heavy-light form factors. For a heavy final meson $B\to D^{(*)}\ell\nu$,
the normalization is fixed from the symmetry, with the leading power
corrections of order $\Lambda/m_b$ vanishing for certain form factors \cite{Luke}.
No such information is available for light final mesons, 
although several properties of the heavy-to-light form factors can be
established in a model-independent way.

The heavy mass scaling of the form factors can be straightforwardly derived 
from the mass dependence of the $|B\rangle$ states implicit in
their relativistic normalization $|\bar B(p)\rangle \sim \sqrt{m_b}$. 
Adopting the usual definition of the formfactors (see, e.g. \cite{BeFe}), 
one finds the scaling laws \cite{IsWi}
\bea\label{pc}
& &T_1 + \frac{m_B^2-m_V^2}{q^2} (T_1-T_2) \propto m_b^{1/2}\,,\qquad
T_1 - \frac{m_B^2-m_V^2}{q^2} (T_1-T_2) \propto m_b^{-1/2}\\
& &\hspace{3.3cm} V(q^2) \propto m_b^{1/2}\,,\qquad 
A_1(q^2) \propto m_b^{-1/2}\,.\nonumber
\eea 
Relations among form factors of different currents can also be derived.
There are three such relations for a vector light meson, and one relation 
for the pseudoscalar meson. For example, two of the $B\to V$ relations are \cite{IsWi}
\bea\label{hqs1}
T_1 + \frac{m_B^2-m_V^2}{q^2} (T_1-T_2) &=& \frac{2m_B}{m_B+m_V} V(q^2) +
O(m_b^{-1/2})\\
\label{hqs2}
T_1 - \frac{m_B^2-m_V^2}{q^2} (T_1-T_2) &=& 
-\frac{2E}{m_B+m_V} V(q^2) + \frac{m_B+m_V}{m_B} A_1(q^2) +
O(m_b^{-3/2})\,.
\eea
These relations are relevant for a method discussed in
Refs.~\cite{SY,Vub} for determining the CKM matrix element $|V_{ub}|$ from exclusive
$B$ decays. This method combines data on
semileptonic $B\to \rho\ell \nu$ and rare radiative decays $B\to K^* \ell^+
\ell^-$ near the zero recoil point, and $|V_{ub}|$ is extracted from the ratio
\cite{SY,Vub}
\bea\label{vub}
\frac{\mbox{d}\Gamma(B\to \rho e\nu)/\mbox{d}q^2}
{\mbox{d}\Gamma(B\to K^* \ell^+\ell^-)/\mbox{d}q^2} =
\frac{8\pi^2}{\alpha^2} 
\frac{|V_{ub}|^2}{|V_{tb} V^*_{ts}|^2} 
\frac{1}{|C_9|^2+|C_{10}|^2} 
\frac{|A^{B\to \rho}_1(q^2)|^2}{|A_1^{B\to K^*}(q^2)|^2} 
\frac{(m_B+m_\rho)^2}{(m_B+m_{K^*})^2}
\frac{1}{1+\Delta(q^2)}
\eea
The SU(3) breaking in this 
result can be eliminated using a Grinstein-type double ratio \cite{ben} and data on 
semileptonic $D\to K^*(\rho) e\bar\nu$ decays, resulting in a $|V_{ub}|$ 
determination at the 10\% level \cite{Vub}.

Recently, the leading power corrections to the heavy quark symmetry relations 
Eqs.~(\ref{hqs1}), (\ref{hqs2}) have been computed in Ref.~\cite{GP1}. 
Contrary to naive expectations, they have a very simple form and depend only 
on the form factors of the dimension-4 currents $\bar q iD_\mu (\gamma_5) b$.
These corrections are required for example to determine the tensor form factor
$T_1(q^2)$ in terms of $V,A_1$ measured in exclusive semileptonic
$B\to V\ell \nu$ decays. It is easy to see that
combining the symmetry relations Eqs.~(\ref{hqs1}), (\ref{hqs2}) in order to 
extract $T_1$ is possible only if the leading correction of $O(m_b^{-1/2})$ to
Eq.~(\ref{hqs1}) is known (since the latter is of the same order as the
terms shown on the RHS of Eq.~(\ref{hqs2})). 

We give here an alternative derivation of this relation, and generalize it beyond
the low recoil assumption implicit in the HQET derivation in \cite{GP1}.
Using the QCD equation of motion for the quark fields one finds
\bea
i\partial^\nu (\bar q i\sigma_{\mu\nu} b) = -(m_b+m_q) \bar q\gamma_\mu
b + 2\bar qiD_\mu b - i\partial_\mu(\bar q b)\,.
\eea
Taking the $B\to V$ matrix element of this relation one finds
\bea
T_1(q^2) = \frac{m_b+m_q}{m_B+m_V} V(q^2) - {\cal D}(q^2) \to
\left\{
\begin{array}{ll}
\frac{\displaystyle m_B-\bar\Lambda}{\displaystyle m_B+m_V} V(q^2) - {\cal D}(q^2) +
O(m_b^{-3/2})\quad \mbox{(low recoil)}\\
V(q^2) - {\cal D}(q^2) + O(Q^{-5/2})\hspace{2.0cm} \mbox{(large energy)}\\
\end{array}
\right.  
\eea 
The first equality is exact and holds for arbitrary recoil, while
the second relation gives its asymptotic form in the low recoil and large energy
regions, respectively.  The form factor ${\cal D}(q^2)$ (scaling like ${\cal D}
\propto m_b^{-1/2}$ in the low recoil region) is defined as 
\bea
 \langle V(p',\varepsilon)|\bar q\, iD_\mu b| \bar B(p)\rangle &=& i{\cal D}(q^2)
\epsilon_{\mu\nu\rho\sigma} \varepsilon^{*\nu} p^{\rho} p^{\prime\sigma} \,,
\eea 
and vanishes exactly in the constituent quark model \cite{GP1}, which
suggests that its value could be small. It would be interesting to see if this
suppression is confirmed by nonperturbative methods such as QCD sum rules or
lattice QCD.

Similar results are obtained in Ref.~\cite{GP1} for subleading power 
corrections to all the other $B\to P,V$ form factor relations in the zero-recoil 
region. In all these cases the subleading terms depend only on
form factors of the local dimension-4 operators $\bar q iD_\mu (\gamma_5) b$.
These results were used in Ref.~\cite{GP2} to estimate the subleading corrections of 
$O(\Lambda/m_b)$ to the $|V_{ub}|$ determination using Eq.~(\ref{vub}).
These corrections can be as large as 5\%, and are dominated by one of the (unknown) 
form factors of $\bar q iD_\mu \gamma_5 b$. 
Quark model estimates of this matrix element suggest that the correction is under
a few percent, and more precise determinations (lattice QCD) could help to
reduce or eliminate this source of uncertainty.
\vspace{0.2cm}

{\em Note added.} The convergence of the $k_+$ convolution in Eq.~(\ref{fact})
was shown in \cite{LaNe}, and
very recently further work on heavy-light form factors
within SCET studying the convergence of the convolution integrals was reported
in \cite{BFtalk}.

\section*{Acknowledgments}
The work of D.P. was supported by the DOE under Grant No. DOE-FG03-97ER40546 
and by the US National Science Foundation Grant PHY-9970781, and I.S. was
supported by the U.S. Department of Energy (DOE) under the cooperative research
agreement DF-FC02-94ER40818.

\section*{Appendix}

We present here the complete results for the form factors in factorized
form.
Adopting the usual parameterization of the form factors, the
explicit results for the $B\to P$ form factors are \cite{ps1}
(we use here the notations of \cite{ps1} for the Wilson coefficients
of SCET operators)
\bea\label{f+}
& &f_+(q^2) = \left( C_1^{(v)} + \frac{E}{m_B} C_2^{(v)} + C_3^{(v)}\right)
\zeta^P \\
& & \hspace{2cm} + N_0 \int dx dl_+ \left\{
\frac{2E-m_B}{m_B} \left[
B_1^{(v)} - \frac{E}{m_B-2E} B_2^{(v)} - \frac{m_B}{m_B-2E} B_3^{(v)} \right]
J_a \right.\nonumber \\
& &\hspace{3cm}  +\left. \frac{2E}{m_b}\left[
B_{11}^{(v)} - \frac{E}{m_B} B_{12}^{(v)} - B_{13}^{(v)} \right] J_b\right\}
\phi_\pi(x) \phi_B^+(l_+)\nonumber\\
\label{f0}
& & \frac{m_B}{2E}f_0(q^2) = 
\left( C_1^{(v)} + \frac{m_B-E}{m_B} C_2^{(v)} + C_3^{(v)}\right)
\zeta^P\\
& & \hspace{2cm} + N_0 \int dx dl_+ \left\{
\frac{m_B-2E}{m_B} \left[
B_1 + \frac{m_B-E}{m_B-2E} B_2^{(v)} + \frac{m_B}{m_B-2E} B_3^{(v)} 
\right] J_a \right.\nonumber\\
& &\hspace{3cm} + \left.\frac{2E}{m_b} \left[
B_{11}^{(v)} - \frac{m_B-E}{m_B} B_{12}^{(v)} - B_{13}^{(v)} \right]
J_b\right\} \phi_\pi(x) \phi_B^+(l_+)\nonumber\\
\label{fT}
& &\frac{m_B}{m_B+m_P}
f_T(q^2) =  \left( C_1^{(t)} - C_2^{(t)} - C_4^{(t)}\right)
\zeta^P \\ 
& & + N_0  \int_0^1 dx dl_+ \left\{
-\left[
B_1^{(t)} - B_2^{(t)} - 2B_3^{(t)} + B_4^{(t)}\right] J_a 
- \frac{2E}{m_b} [B_{15}^{(t)} + B_{16}^{(t)} - B_{18}^{(t)}] J_b \right] 
\phi_B^+(l_+) \phi(x)
\nonumber\,,
\eea
with $N_0 = f_B f_P m_B/(4E^2)$.

The corresponding results for the $B\to V$ form factors read
\bea
& &\frac{m_B}{m_B+m_V} V(q^2) =  C_1^{(v)} \zeta^V_\perp 
-  N_\perp \int_0^1 dx dl_+ \left[
-\frac{1}{2} B_4^{(v)} J_{a}^{\perp} + 
\frac{E}{m_b} (2B_{11}^{(v)} + B_{14}^{(v)}) J_{b}^{\perp} \right]
\phi_B^+(l_+) \phi_\perp(x) \nonumber \\
& &\frac{m_B+m_V}{2E}
 A_1(q^2) =  C_1^{(a)} \zeta^V_\perp 
- N_\perp \int_0^1 dx dl_+ \left[
-\frac{1}{2} B_4^{(a)} J_{a}^{\perp} +
\frac{E}{m_b} (2B_{11}^{(a)} + B_{14}^{(a)}) J_{b}^{\perp} \right] 
\phi_B^+(l_+)\phi_\perp(x)  \nonumber\\
& &A_0(q^2) = \left( C_1^{(a)} + \frac{m_B-E}{m_B} C_2^{(a)}
+ C_3^{(a)} \right) \zeta^V_\parallel \\
& &  \hspace{2cm} + N_\parallel 
\int_0^1 dx dl_+ \left\{\left[ 
\frac{m_B-2E}{m_B} B_1^{(a)} + \frac{m_B-E}{m_B} B_2^{(a)} + B_3^{(a)} \right]
J_a \right. \nonumber\\
& & \hspace{3cm} - \left. \frac{2E}{m_b} \left[ -B_{11}^{(a)} +
\frac{m_B-E}{m_B} B_{12}^{(a)} + B_{13}^{(a)} \right] J_b \right\}
\phi_B^+(l_+) \phi_\parallel(x)  \nonumber \\
& &\frac{m_B E}{m_B+m_V}
A_2(q^2) - \frac12 (m_B+m_V) A_1(q^2) = - 
\left( C_1^{(a)} + \frac{E}{m_B} C_2^{(a)} + C_3^{(a)}  \right) m_V 
\zeta^V_\parallel \\
& &\hspace{2cm}  + m_V N_\parallel 
\int_0^1 dx dl_+ \left\{\left[ 
\frac{m_B-2E}{m_B} B_1^{(a)} - \frac{E}{m_B} B_2^{(a)} - B_3^{(a)} \right]
J_a \right. \nonumber\\
& & \hspace{3cm} -\left. \frac{2E}{m_b} 
\left[ B_{11}^{(a)} - \frac{E}{m_B} B_{12}^{(a)} - B_{13}^{(a)} 
\right] J_b \right\}
\phi_B^+(l_+) \phi_\parallel(x)  \nonumber \\
\label{T1}
& &T_1(q^2) = \frac{m_B}{2E} T_2(q^2) = \left\{
C_1^{(t)} - \frac{m_B-E}{m_B} C_2^{(t)} - C_3^{(t)} \right\} \zeta^V_\perp\\
& & \hspace{2cm} - \frac12 N_\perp 
\int_0^1 dx dl_+ \left\{ \left[ B_5^{(t)} + \frac{m_B-E}{m_B} B_6^{(t)} \right]
J_a^{\perp} \right.\nonumber\\
& & \hspace{1cm} -\left. \frac{2E}{m_b} \left[
2B_{15}^{(t)} + 2B_{17}^{(t)} + B_{19}^{(t)} + B_{21}^{(t)} + 
\frac{m_B-E}{m_B}(2B_{16}^{(t)} + B_{20}^{(t)})
\right] J_b^{\perp} \right\} \phi_B^+(l_+) \phi_\perp(x)\nonumber\\
& & E T_3(q^2) - \frac{m_B}{2}T_2(q^2) = 
- (C_1^{(t)} - C_2^{(t)} - C_4^{(t)}) m_V \zeta^V_\parallel \\
& & \hspace{0.5cm} + m_V N_\parallel 
\int_0^1 dx dl_+ \left\{ \left[ B_1^{(t)} -  B_2^{(t)} -
2B_3^{(t)} - B_4^{(t)} \right]J_a 
 +  \frac{2E}{m_b} (B_{15}^{(t)} + B_{16}^{(t)} - B_{18}^{(t)} )
J_b \right\}
\phi_B^+(l_+) \phi_\parallel(x)\nonumber
\eea
where $N_\perp = m_B/(4E^2) f_B f_V^T$ and $N_\parallel = m_B/(4E^2) f_B f_V$.
There are 2 jet functions $J_{a,b}$ contributing to $B\to P, V_\parallel$ 
(defined as in \cite{ps1}), and 2 other jet functions contributing only to 
$B\to V_\perp$, denoted as $J_{a,b}^{\perp}$. At tree level they are equal 
$J_{a,b}^{(\perp)}(z,x,l_+) = \frac{\pi\alpha_s C_F}{N_c} \frac{1}{\bar x l_+}$, but in
general they are different.
The Wilson coefficients satisfy $C_{1-3}^{(v)} = C_{1-3}^{(a)}$ and 
$B_{1-4}^{(v)} = B_{1-4}^{(a)}$ in the NDR scheme.
Reparameterization invariance constrains them as $B_{1-3}^{(v,a,t)} =
C_{1-3}^{(v,a,t)}$, $B_4^{(v,a)} = -2C_3^{(v,a)}$, $B_4^{(t)} = C_4^{(t)}$,
$B_5^{(t)} = 2C_3^{(t)}$, $B_6^{(t)} = -2C_4^{(t)}$ \cite{rpi,ps1}.
At tree level they are given by $C_1^{(v,a,t)}=1$, $B_1^{(v,a,t)}=1$, 
$B_{13}^{(v,a)}=-1$,
$B_{17}^{(t)} = 1$.


\begin{thebibliography}{99}



\bibitem{rare}
A.~Ali, P.~Ball, L.~T.~Handoko and G.~Hiller,
Phys.\ Rev.\ D {\bf 61}, 074024 (2000);
A.~Ali, E.~Lunghi, C.~Greub and G.~Hiller,
Phys.\ Rev.\ D {\bf 66}, 034002 (2002);
A.~Ali and E.~Lunghi,
Eur.\ Phys.\ J.\ C {\bf 21}, 683 (2001);
G. Hiller, {\em Phenomenology of new physics},
in these proceedings.




\bibitem{exp} 
B.~Aubert {\it et al.}  [BABAR Collaboration],
arXiv:hep-ex/0207080;
S.~B.~Athar {\it et al.}  [CLEO Collaboration],
arXiv:hep-ex/0304019;
B.~Aubert {\it et al.}  [BABAR Collaboration],
arXiv:hep-ex/0308042;
K.~Abe  [Belle Collaboration],
arXiv:hep-ex/0308044;
E. Thorndike, {\em Inclusive and exclusive $V_{ub}$ measurements},
in these proceedings;
A. Ishikawa, {\em Review of $b\to s \ell^+ \ell^-$ and $B\to \ell^+ \ell^-$ decays},
in these proceedings.


\bibitem{pball}  
P.~Ball,
arXiv:hep-ph/0306251;
P.~Ball,
arXiv:hep-ph/0308249.



\bibitem{becirevic-fpcp03} D. Becirevic, 
{\em Lattice QCD for B decays},
in these proceedings.


\bibitem{scet}
C.~W.~Bauer, S.~Fleming and M.~E.~Luke,
Phys.\ Rev.\ D {\bf 63}, 014006 (2001);
C.~W.~Bauer, S.~Fleming, D.~Pirjol and I.~W.~Stewart,
Phys.\ Rev.\ D {\bf 63}, 114020 (2001);
C.~W.~Bauer and I.~W.~Stewart,
Phys.\ Lett.\ B {\bf 516}, 134 (2001).

\bibitem{fleming-fpcp03} S. Fleming, 
{\em The Soft Collinear Effective Theory},
in these proceedings.


\bibitem{bpssoft}
C.~W.~Bauer, D.~Pirjol and I.~W.~Stewart,
Phys.\ Rev.\ D {\bf 65}, 054022 (2002)

\bibitem{rpi}
J.~Chay and C.~Kim,
Phys.\ Rev.\ D {\bf 65}, 114016 (2002);
A.~V.~Manohar, T.~Mehen, D.~Pirjol and I.~W.~Stewart,
Phys.\ Lett.\ B {\bf 539}, 59 (2002).

\bibitem{bcdf}
M.~Beneke, A.~P.~Chapovsky, M.~Diehl and T.~Feldmann,
Nucl.\ Phys.\ B {\bf 643}, 431 (2002);
M.~Beneke and T.~Feldmann,
Phys.\ Lett.\ B {\bf 553}, 267 (2003).


\bibitem{ps1}
D.~Pirjol and I.~W.~Stewart,
Phys.\ Rev.\ D {\bf 67}, 094005 (2003).



\bibitem{lpw}
E.~Lunghi, D.~Pirjol and D.~Wyler,
Nucl.\ Phys.\ B {\bf 649}, 349 (2003).

\bibitem{b2gpapers} 
G.~P.~Korchemsky, D.~Pirjol and T.~M.~Yan,
Phys.\ Rev.\ D {\bf 61}, 114510 (2000);
S.~Descotes-Genon and C.~T.~Sachrajda,
Nucl.\ Phys.\ B {\bf 650}, 356 (2003)
S.~W.~Bosch, R.~J.~Hill, B.~O.~Lange and M.~Neubert,
Phys.\ Rev.\ D {\bf 67}, 094014 (2003).


\bibitem{DS} 
S.~Descotes-Genon and C.~T.~Sachrajda,
Phys.\ Lett.\ B {\bf 557}, 213 (2003).



\bibitem{BL} S.~J. Brodsky and P. Lepage, in {\em Perturbative Quantum
Chromodynamics} (World Scientific, Singapore, 1989), pp. 93-240.

\bibitem{akhoury}
R.~Akhoury, G.~Sterman and Y.~P.~Yao,
Phys.\ Rev.\ D {\bf 50}, 358 (1994).


\bibitem{BeFe}
M.~Beneke and T.~Feldmann,
Nucl.\ Phys.\ B {\bf 592}, 3 (2001)

\bibitem{bps5} 
C.~W.~Bauer, D.~Pirjol and I.~W.~Stewart,
Phys.\ Rev.\ D {\bf 67}, 071502 (2003).

\bibitem{li}   
T.~Kurimoto, H.~n.~Li and A.~I.~Sanda,
Phys.\ Rev.\ D {\bf 65}, 014007 (2002).

\bibitem{lireview}
H.~n.~Li,
arXiv:hep-ph/0303116.


\bibitem{gauge}
C.~W.~Bauer, D.~Pirjol and I.~W.~Stewart,
arXiv:hep-ph/0303156.

\bibitem{Charles} J. Charles et al., Phys. Rev. D {\bf 60}, 014001 (1999).


\bibitem{sudakov}
S.~Descotes-Genon and C.~T.~Sachrajda,
Nucl.\ Phys.\ B {\bf 625}, 239 (2002).


\bibitem{HiNe}
R.~J.~Hill and M.~Neubert,
Nucl.\ Phys.\ B {\bf 657}, 229 (2003)




\bibitem{4q}
A.~J.~Buras, A.~Czarnecki, M.~Misiak and J.~Urban,
Nucl.\ Phys.\ B {\bf 631}, 219 (2002);
M.~Beneke, T.~Feldmann and D.~Seidel,
Nucl.\ Phys.\ B {\bf 612}, 25 (2001);
S.~W.~Bosch and G.~Buchalla,
Nucl.\ Phys.\ B {\bf 621}, 459 (2002);
A.~Ali and A.~Y.~Parkhomenko,
Eur.\ Phys.\ J.\ C {\bf 23}, 89 (2002);
J.~g.~Chay and C.~Kim,
arXiv:hep-ph/0305033.


\bibitem{wa} 
B.~Grinstein and D.~Pirjol,
Phys.\ Rev.\ D {\bf 62}, 093002 (2000);
A.~Hardmeier, E.~Lunghi, D.~Pirjol and D.~Wyler,
arXiv:hep-ph/0307171.


\bibitem{IsWi} 
N.~Isgur and M.~B.~Wise,
Phys.\ Rev.\ D {\bf 42}, 2388 (1990).


\bibitem{Luke}
N.~Isgur and M.~B.~Wise,
Phys.\ Lett.\ B {\bf 237}, 527 (1990);
M.~E.~Luke,
Phys.\ Lett.\ B {\bf 252}, 447 (1990).

\bibitem{SY} A.~I.~Sanda and A.~Yamada,
Phys.\ Rev.\ Lett.\  {\bf 75}, 2807 (1995).



\bibitem{Vub}
Z.~Ligeti and M.~B.~Wise,
Phys.\ Rev.\ D {\bf 53}, 4937 (1996);
Z.~Ligeti, I.~W.~Stewart and M.~B.~Wise,
Phys.\ Lett.\ B {\bf 420}, 359 (1998).


\bibitem{ben}
B.~Grinstein,
Phys.\ Rev.\ Lett.\  {\bf 71}, 3067 (1993).



\bibitem{GP1}
B.~Grinstein and D.~Pirjol,
Phys.\ Lett.\ B {\bf 533}, 8 (2002)

\bibitem{GP2}
B.~Grinstein and D.~Pirjol,
Phys.\ Lett.\ B {\bf 549}, 314 (2002).

\bibitem{LaNe}
B.~O.~Lange and M.~Neubert,
arXiv:hep-ph/0303082.


\bibitem{BFtalk}
M. Beneke and Th. Feldmann, CERN report CERN-TH/2003-202, hep-ph/0308303.


\end{thebibliography}
\end{document}